\documentclass[journal,transmag]{IEEEtran}

\usepackage{graphicx}
\usepackage{amsmath}

\ifCLASSINFOpdf
\else
\fi
\begin{document}
\title{Voltage Uncertainties in the Presence of Photovoltaic Systems }

\author{\IEEEauthorblockN{Katherine Hughes\IEEEauthorrefmark{1}
}
\IEEEauthorblockA{\IEEEauthorrefmark{1}Hawaii Data Analytics LLP
}}

\IEEEtitleabstractindextext{%
\begin{abstract}

With the rising demand for solar energy installation, there is a pressing need for utilities to regulate the voltages at the distribution level. In grids with high penetration of photovoltaic (PV) systems, voltage fluctuations can occur at the distribution systems, resulting in inverter tripping and insufficient power to meet the load. We present a linear model for voltage rise versus PV output power. This model can be used to study the effect of increasing PV system capacities on distribution system voltages. It is observed that voltage fluctuations have greater correlation with the location of the PV systems on the grid than with the PV system capacities, i.e., more randomness and disorder in the behavior of voltage occurs with PV systems with larger line impedance.

\end{abstract}

}

\maketitle

\IEEEdisplaynontitleabstractindextext

\IEEEpeerreviewmaketitle

\section{Introduction}

\IEEEPARstart{N}{ew} methods of mitigating voltage violations in the penetration of distributed generation \cite{c15}-\cite{c17} may allow renewable energies to enter the grid without causing undue harm to existing systems. When an inverter injects power to the grid, the impedance from the grid and inverter output-circuit, cause a voltage rise at the point of common coupling (PCC). The effect of the voltage rise is considered in the design of PV systems, e.g., the inverter disconnects from the grid if the inverter output voltage exceeds its operating limit. Once the inverter ceases to inject current to the grid, it must monitor the voltage for a while to reconnect to the grid \cite{c6}. The grid voltage can be adjusted to avoid unnecessary inverter tripping and system losses caused by voltage rise, and may lead to cascading events(cite). There is a need to model the voltage rise caused by PV systems to develop an optimal control algorithm to control the transformer Load Tap Changer (LTC), the switched capacitors, the bidirectional step regulators, and the operating voltage range of the inverters. Traditional analyses presume that loads exhibit linear elements, i.e. can be represented by constant impedances, on the contrary they may exhibit voltage dependent characteristics.

Switching from grid operator distribution, to consumer supplies of power causes difficulties in power flow. For instance, reverse power flow at low voltage distribution grids can cause a voltage rise and, similarly, if generation exceeds the load, the voltage increases. The liklihood of voltage rise increases with PV systems operating at the end of a distribution line with unity power factor \cite{c1}. As such, voltage fluctuations caused by  PV systems, limit their penetration on the grid \cite{c10}.

\cite{c7} prevent unwanted PV shut-off by applying batteries to store excess energy. \cite{c8} propose a closed-loop control system to regulate the PCC voltage of a PV system by controlling the reactive power injected into the grid. They do not consider the dynamic behaviour of loads and other inverters connected to the distribution grid on the performance of the designed PCC voltage controller. In order to regulate the PCC voltage of a PV system, the \emph{dynamic behavior} of the grid should be known. Based on the authors' works, the dynamic effect of loads and PV inverters on the voltage have yet to be studied.

In a grid with high penetration of PV systems, having a flatter voltage is more expensive, as it may require more voltage regulators \cite{c1}. It has been difficult to quantify the adjustment of voltage regulators and capacitance banks in the presence of high penetration of renewable energies, because of the stochastic nature of the loads and renewable energies. Hence, it is necessary to create a stochastic model for voltage variations through the grid, in order to determine the optimal locations in which to position the voltage regulators, and optimal strategies to operate them. The grid model with a high number of small distributed generators can be described more as a dynamic and non-stationary process. Novel methods for joint analysis of multiple data sets (\cite{ashkan}-\cite{ashkan1}) enable us to develop better models of the voltage variations.

Conventional voltage regulators are not designed for the high variability in voltage caused by PV systems. For example high penetration of PV systems can increase the number of tap changes and lead to degradation. A standard time delay is considered to avoid the load tap changer operations, when the voltage is outside the bandwidth for a short duration. To have a reliable grid with high penetration of PV systems, it is necessary to have model-based voltage regulators that respond quickly to voltage variations. Smart inverters can be used to mitigate the voltage variations caused by variability in PV system output and stochastic load \cite{c5}. 
This would require developing a national standard for inverters to autonomously regulate the voltage. It is important to determine the impact of PV systems on the voltage, to develop strategies for the voltage regulators. The main contributions of this work are as follows.

\subsection{Main Contributions}
The regulator tap position is a function of the PV systems' outputs and loads \cite{c9}. The effect of PV output and load on the voltage should be modeled in order to determine the tap position at each time and the number of tap changes. In this work, we study the impact of distributed PV systems on the voltage rise of the distribution lines. We quantify the variability of the voltage at different distances from the transformer. We use partial least squares regression to model voltage variation versus injected power in individual PV systems. The accuracy of the model is analyzed by presenting the statistical parameters of voltage variations. \\

The rest of the paper is structured as follows. In Section~\ref{SectionII} the data that is used in the analysis is described and the numerical results for PV profile analysis is presented. Section~\ref{SectionIV} implements partial least square to predict the voltage rise caused by the inverter output. Finally, concluding remarks are provided in Section~\ref{SectionVII}.

\section{PV Profile}\label{SectionII}
A customer's net energy is defined as the difference between the energy supplied by the grid and energy generated by the customer's PV system. All five of the consumers in this study are equipped with net energy meters that report the net energy and PCC voltage at one minute intervals. All consumers in this study are also equipped with a second meter that reports the PV generation (inverter output) and voltage at one-minute intervals. The reported PV generations are the aggregate of the PV generation over the one-minute intervals. The root mean square (RMS) voltage is measured at each second and the maximum voltage over one-minute intervals is reported. The analysis is based on $160$ days' data of five PV systems with capacities $1.94$, $3.87$, $9.24$, $11.61$ and $7.31$ kilowatt (KW). The PV systems are positioned at different distances from the distribution transformer on the consumers' properties. The line impedance from each PV system to the transformer is given in the Table~\ref{table:tb0}. It is observed that the daily energy for a PV system with capacity of one kilowatt, varies from $3.96$ KWh to $5.17$ KWh. On average, the amount of daily energy for a PV system with capacity of one kilowatt is $4.78$ KWh.

\begin{table}[h]
\caption{Impedance (OHM) and average daily energy per system capacity (SC)}
\label{table:tb0}
\begin{center}
\begin{tabular}{|c||c||c||c||c||c|}
\hline
  Capacity (KW) &1.94&3.87&7.31&11.61&9.24\\
\hline
   Impedance (OHM)  &0.077&0.060&0.053&0.025&0.011\\
\hline
   Average daily energy/SC  &4.67&5.13&3.96&5.17&4.98\\
\hline
\end{tabular}
\end{center}
\end{table}
The efficiency of a PV system can be impacted if part of a PV panel is shaded, or if the panel is positioned at a non-optimal angle or direction. In figure~\ref{fig:pvirrad}, the PV generated power versus irradiance for all the PV systems under consideration is plotted. Each PV system output power in kilowatts (KW) is divided by its PV System Capacity (SC). The vertical axis represents the generated power per one unit of the PV system capacity, and horizontal axis represents the irradiance (kilowatt per square meter). It is observed that the variations of the PV systems output in irradiance can be approximated as a linear model. The scatter plot shows all the sample data points for all systems and the red line shows the best fit for all the systems using linear regression \cite{c4}. Again using linear regression, the PV output power is modeled as a linear function of irradiance $\mathit{P}= 0.906 \,\ \mathit{R}$, in which $\mathit{P}$ is the generated power per unit of the PV system, and $\mathit{R}$ is the irradiance power. The following can be observed from PV data:

\begin{figure}
\begin{center}
\framebox{\parbox{3in}{
\includegraphics[width=8cm]{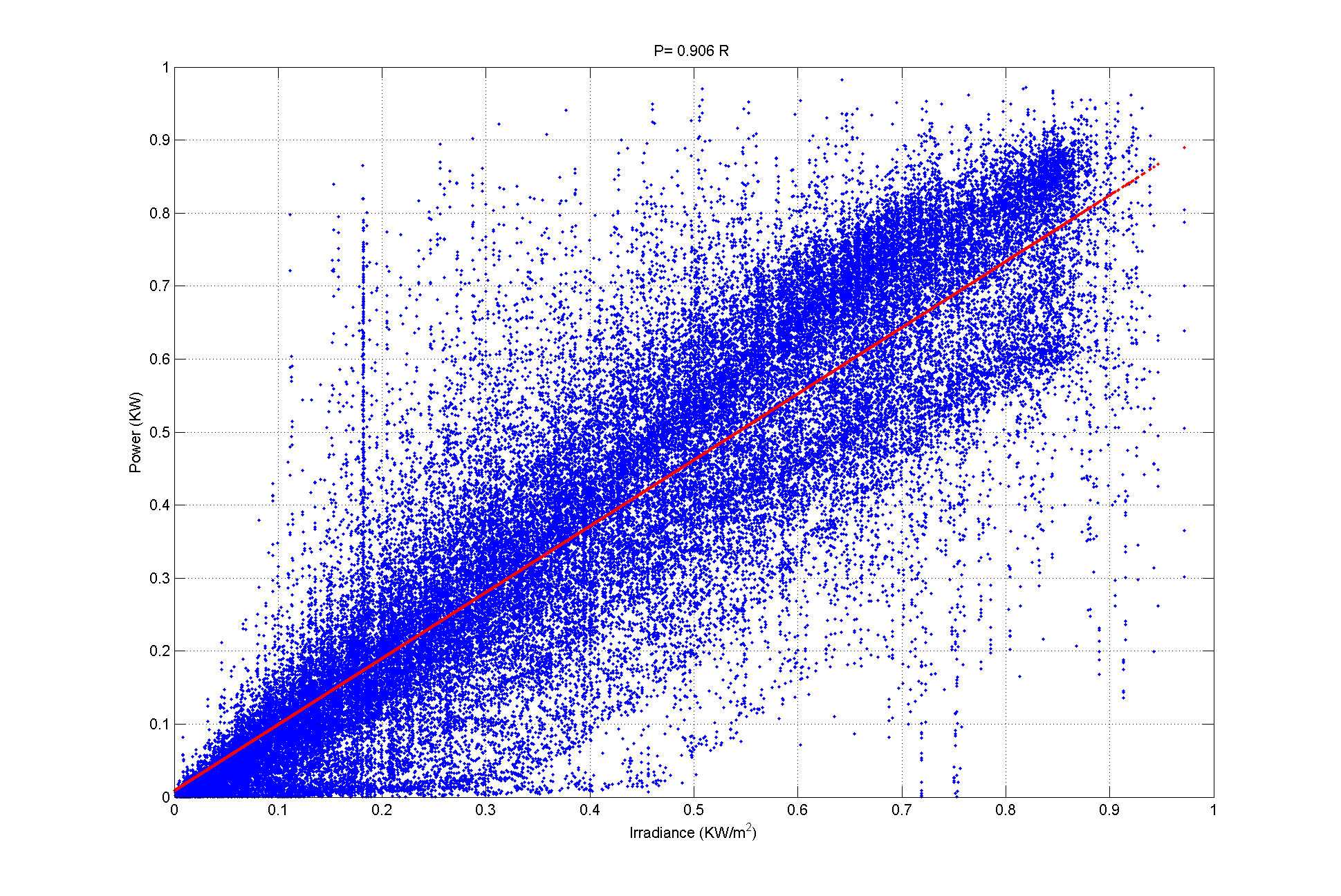}}}
\end{center}
\caption{The vertical axis represents generated power (kilowatt) per unit of PV system capacity, and horizontal axis represents the irradiance (kilowatt per square meter). The blue dots visualize the scatter points of PV systems output versus irradiance. The red line is the best fit to the scatter points using linear regression.}
\label{fig:pvirrad}
\end{figure}
\begin{itemize}
\item The average PV systems' generation is less than $70$ percent of PV system capacity.
\item The efficiency of the PV systems is not related to the capacity of the system.
\item Due to differences in PV panel direction, the peak-time of PV profiles varies. This can be used for the flattening of PV generation in the grid.
\end{itemize}

The effect of irradiance on the PV system output is deterministic on a clear sky day. Because of the irregular presence of the clouds, the irradiance data has been modeled as a random variable \cite{c2}. Voltage is affected by PV system output and the load. PV system output and load are random variables, and consequently the voltage variations in distribution grids can be modeled as a random process. In the next section, the PCC voltage of the five PV systems, each with different line impedance, is analyzed.

\section{Voltage Analysis} \label{SectionIV}

\begin{figure}[h!]
\centering
\framebox{\parbox{3in}{
\includegraphics[width=8cm]{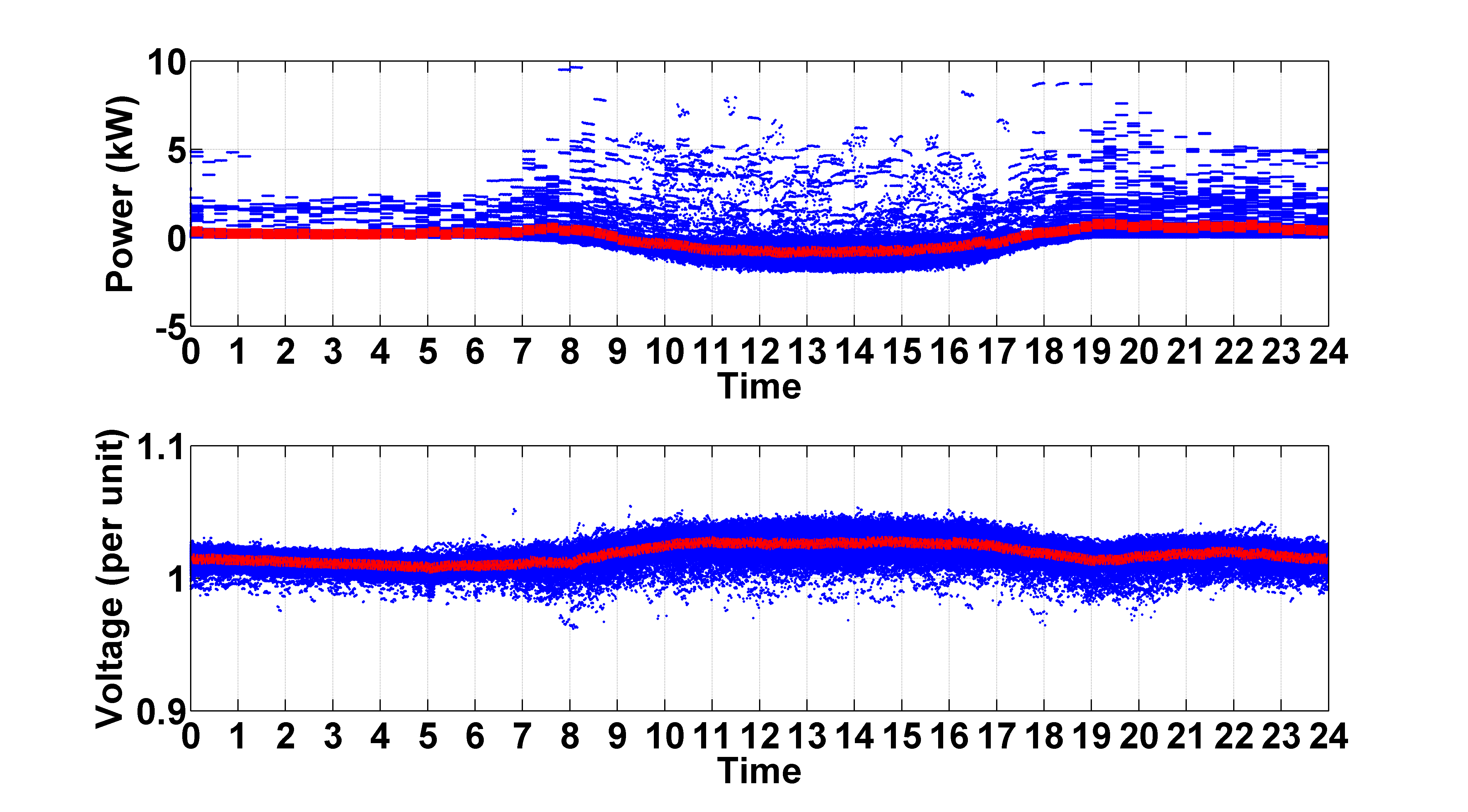}}}
\caption{Power and voltage versus time for the PV system capacity $1.94$ KW .}
\label{fig:povo1}
\framebox{\parbox{3in}{
\includegraphics[width=8cm]{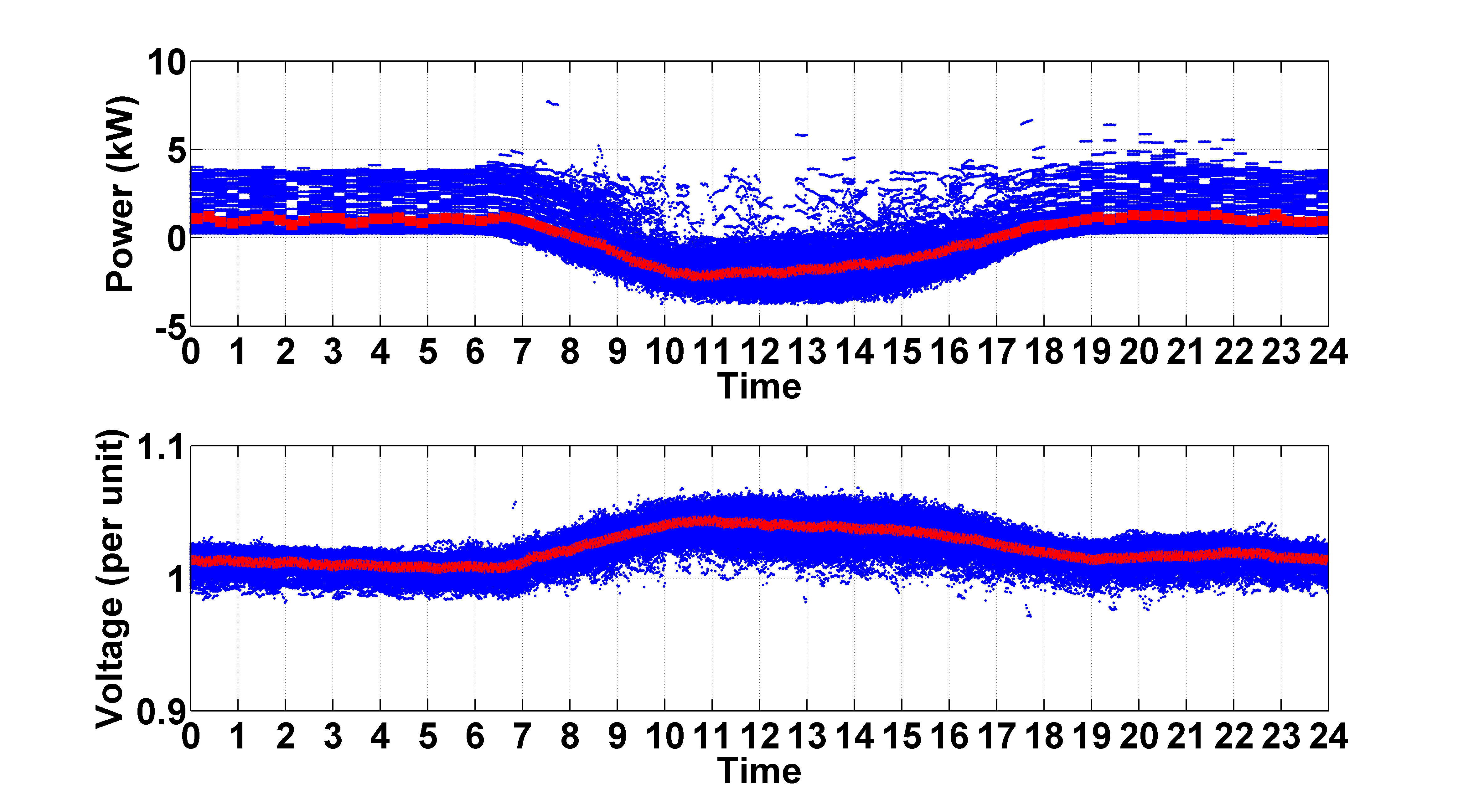}}}
\caption{Power and voltage versus time for the PV system capacity $3.87$ KW.}
\label{fig:povo2}
\framebox{\parbox{3in}{
\includegraphics[width=8cm]{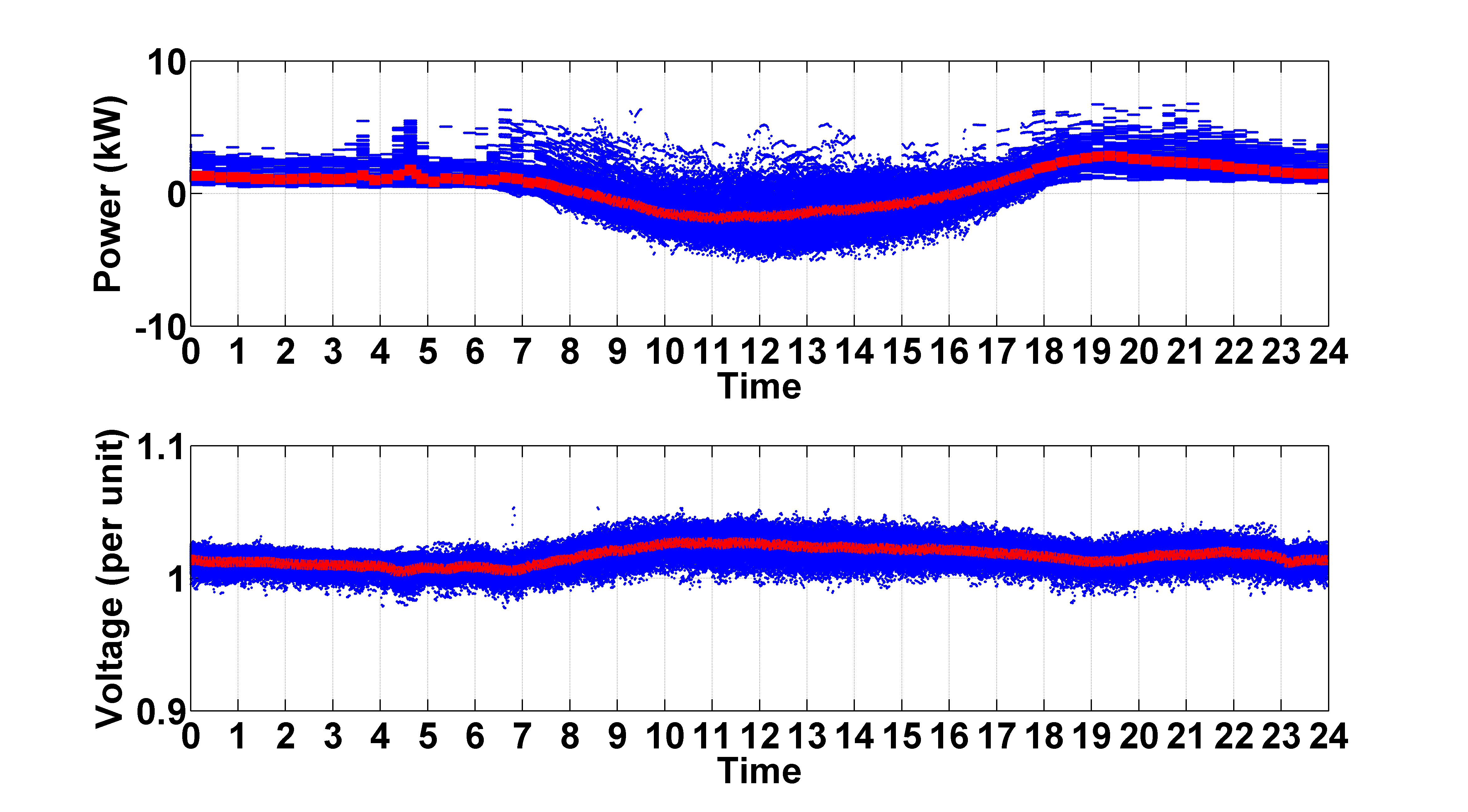}}}
\caption{Power and voltage versus time for the PV system capacity $7.31$ KW. }
\label{fig:povo3}
\end{figure}

\begin{figure}[h!]
\centering
\framebox{\parbox{3in}{
\includegraphics[width=3in]{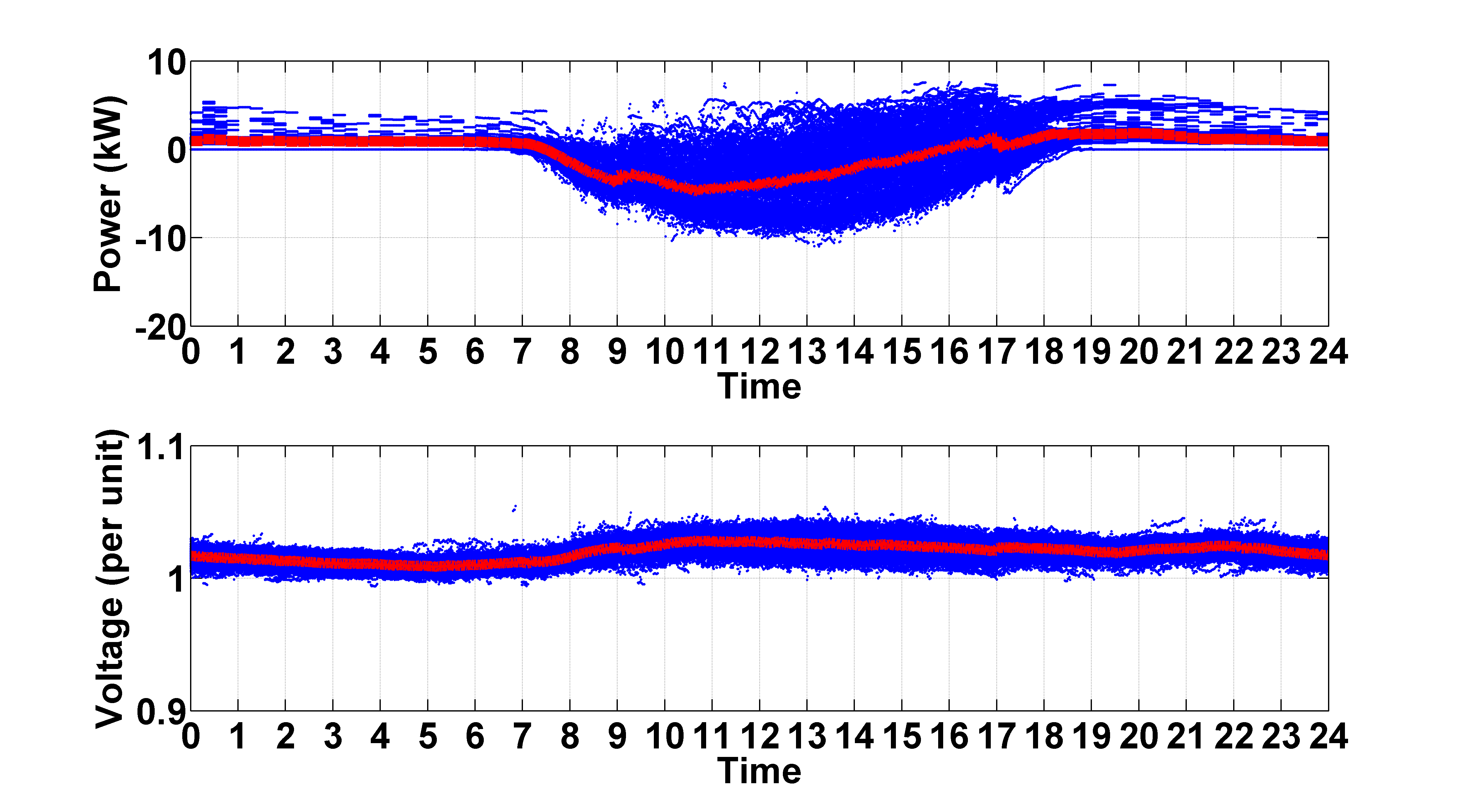}}}
\caption{Power and voltage versus time for the PV system capacity $11.61$ KW }
\label{fig:povo4}
\framebox{\parbox{3in}{
\includegraphics[width=3in]{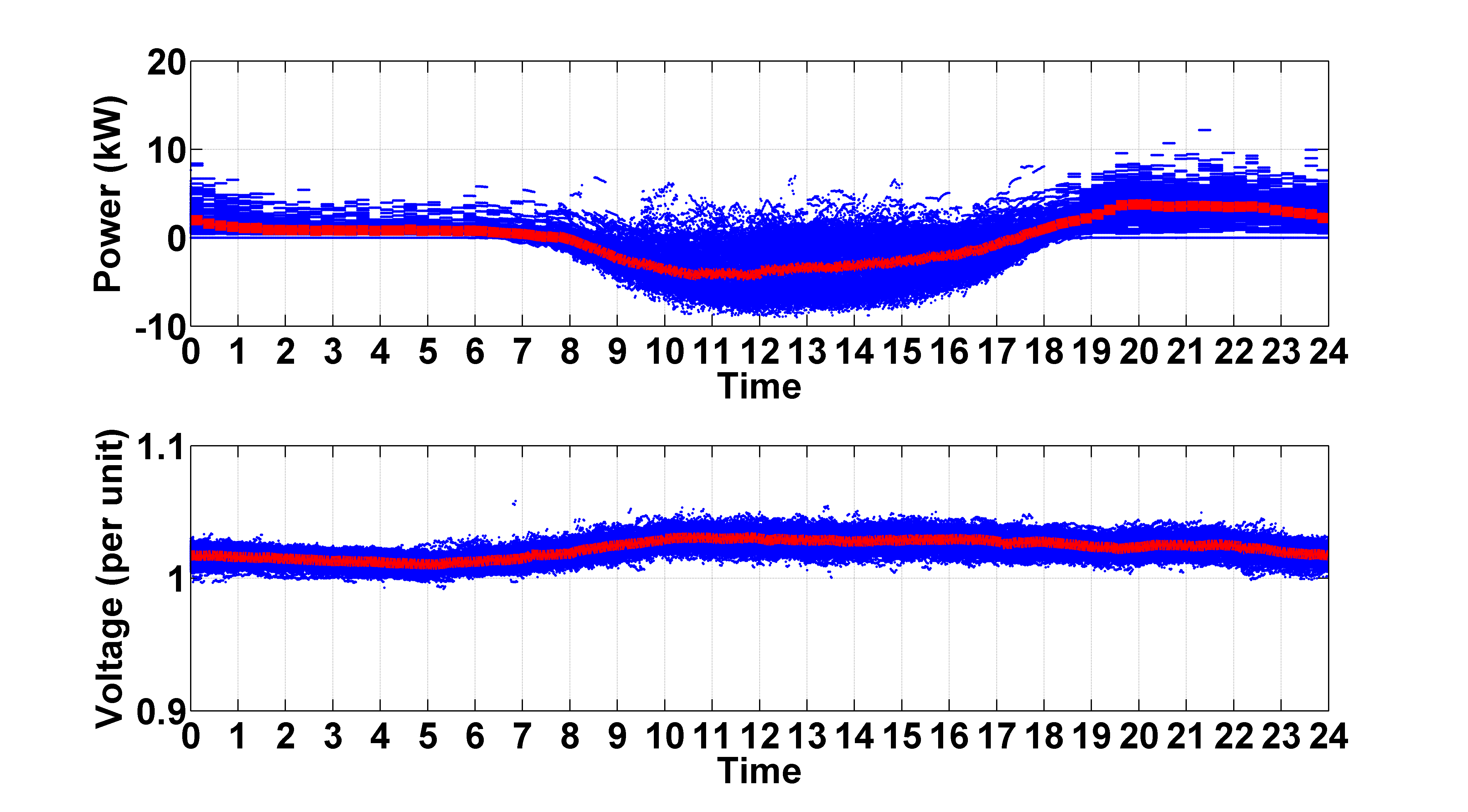}}}
\caption{Power and voltage versus time for the PV system capacity $9.24$ KW. }
\label{fig:povo5}
\end{figure}

In figures~\ref{fig:povo1}-\ref{fig:povo5}, power and voltage versus time are plotted. The blue dots are the scatter points of the data and the red-lines are the averages of the data at each time. Negative power values imply injection of power into the grid and positive power values represent power flowing in the opposite direction from the grid to the house. The PCC voltage of a PV system is dependent on both active and reactive power injected to the grid \cite{c11}-\cite{c14}. In this study, the PV systems only injects active power to the grid, and the amount of reactive load is negligible. Therefore, reactive power is not considered in the voltage analysis. Note that the voltage at each PCC is affected by neighbouring loads, PV generation quantities, and the operation of load tap changers, as well as capacitance banks. The voltage does not rise consistently when injecting more power to the grid at a single PCC, because of the stochastic nature of the load and PV generation at neighboring properties. It is observed that \emph{on average} by injecting more power to the grid, voltage begins to increase. In the next section, the voltage variation is modeled as a function of injected power into the grid, and the statistical parameters about the accuracy of this model are presented.

\subsection{Partial Least Square (PLS)}
\begin{figure}[h!]
\centering
\framebox{\parbox{3in}{
\includegraphics[width=3in]{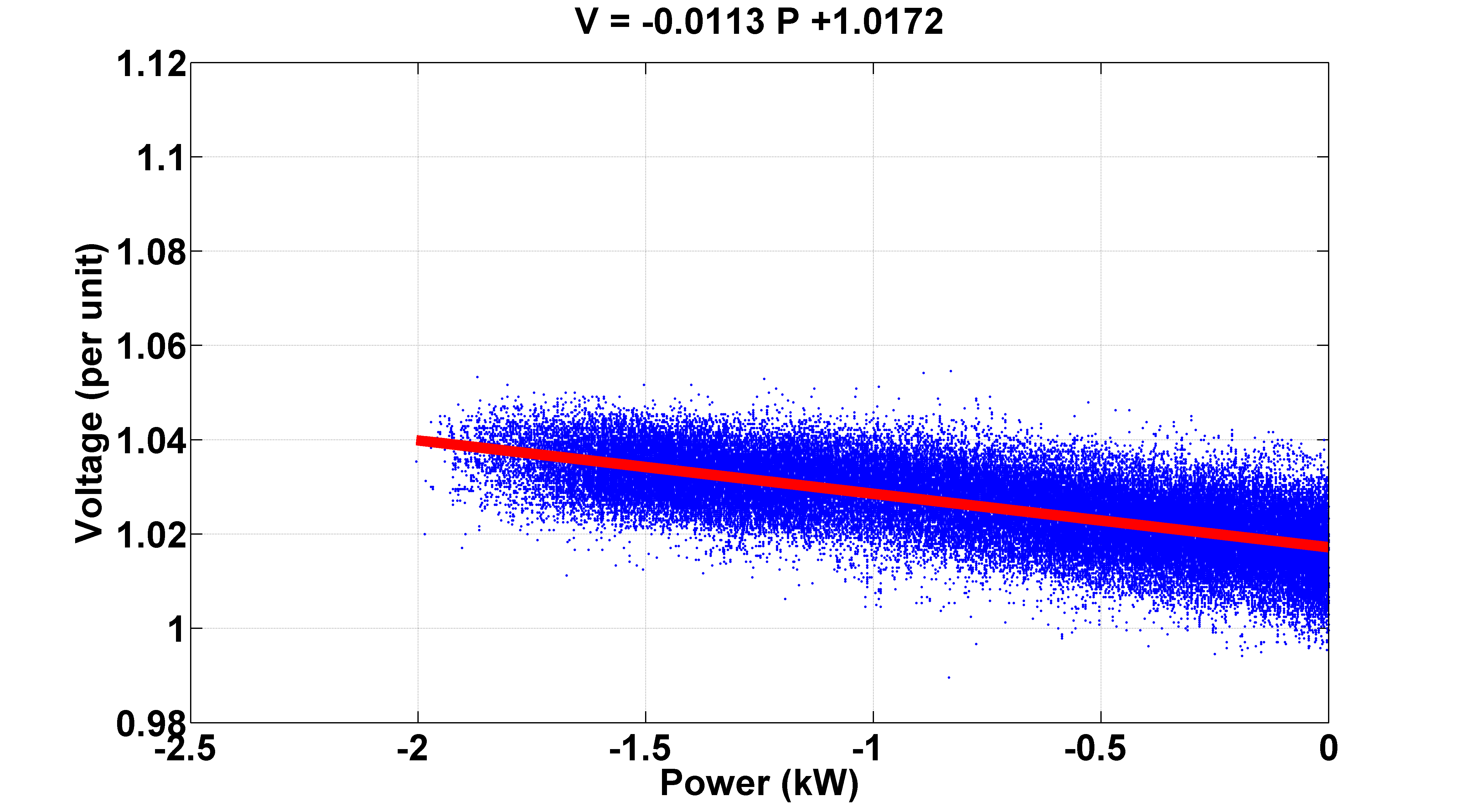}}}
\caption{Voltage versus power for the PV system capacity $1.94$ KW.}
\label{fig:vvp1}
\framebox{\parbox{3in}{
\includegraphics[width=3in]{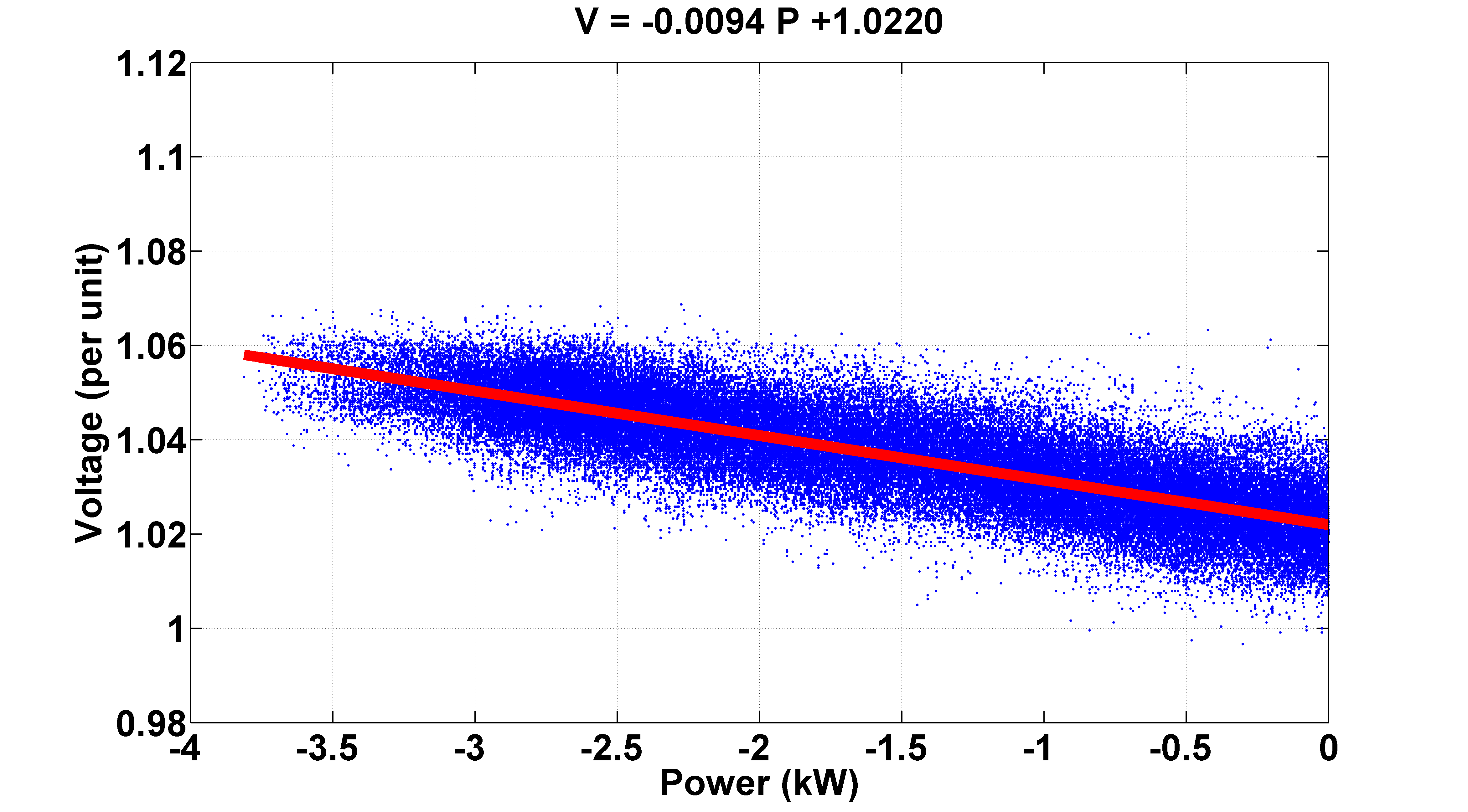}}}
\caption{Voltage versus power for the PV system capacity $3.87$ KW. }
\label{fig:vvp2}
\framebox{\parbox{3in}{
\includegraphics[width=3in]{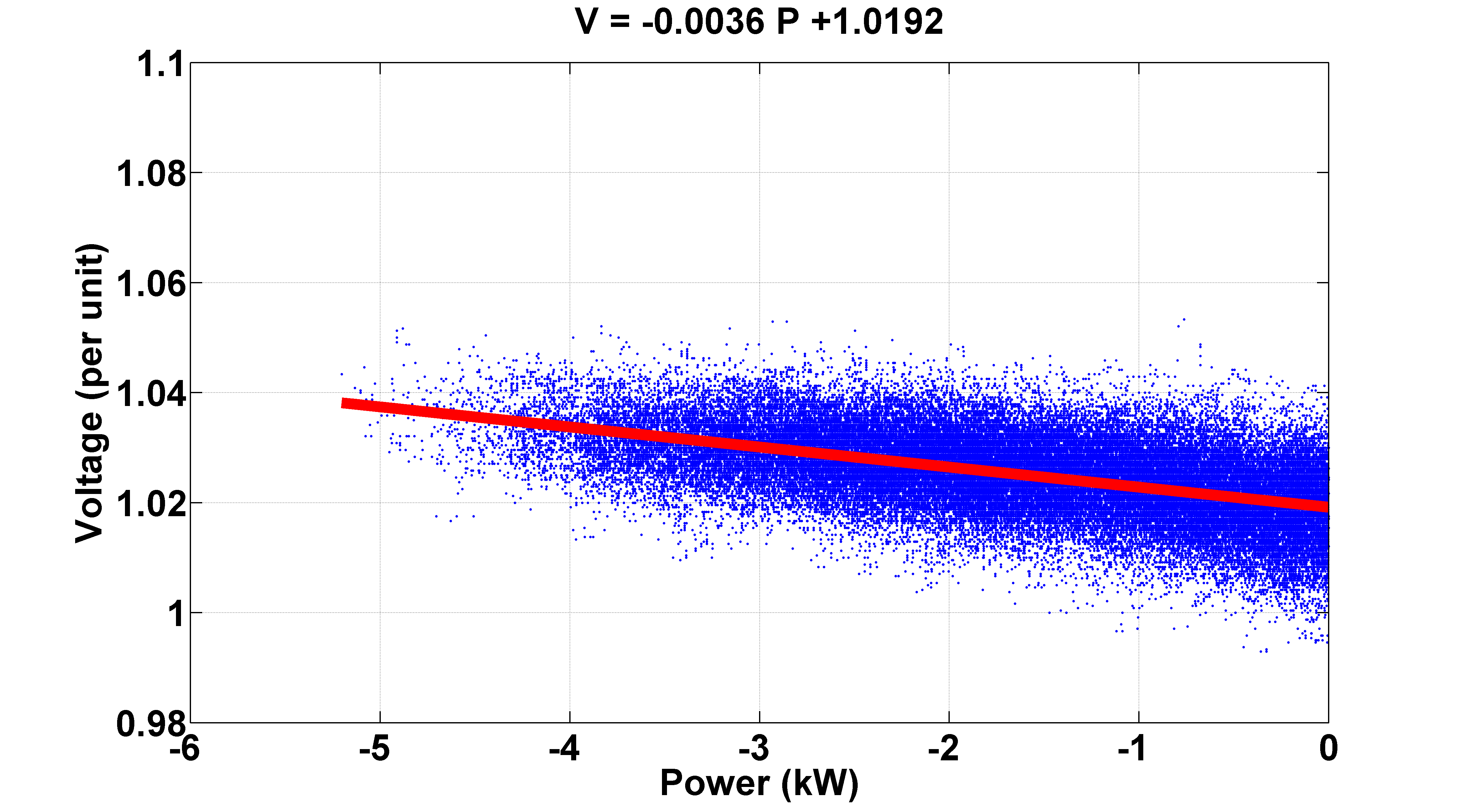}}}
\caption{Voltage versus power for the PV system capacity $7.31$ KW.}
\label{fig:vvp3}
\end{figure}

\begin{figure}[h!]
\centering
\framebox{\parbox{3in}{
\includegraphics[width=3in]{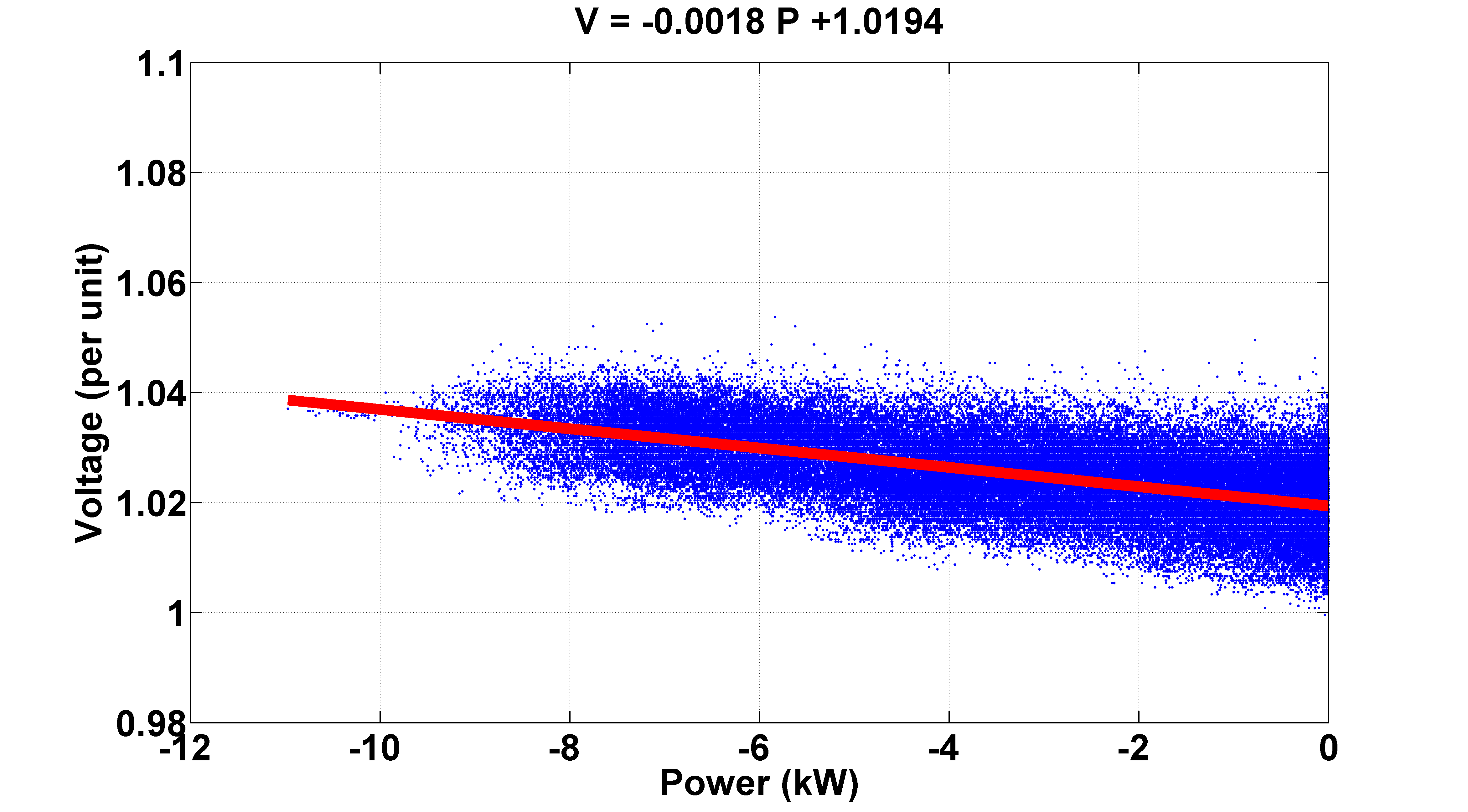}}}
\caption{Voltage versus power for the PV system capacity $11.61$ KW.}
\label{fig:vvp4}
\framebox{\parbox{3in}{
\includegraphics[width=3in]{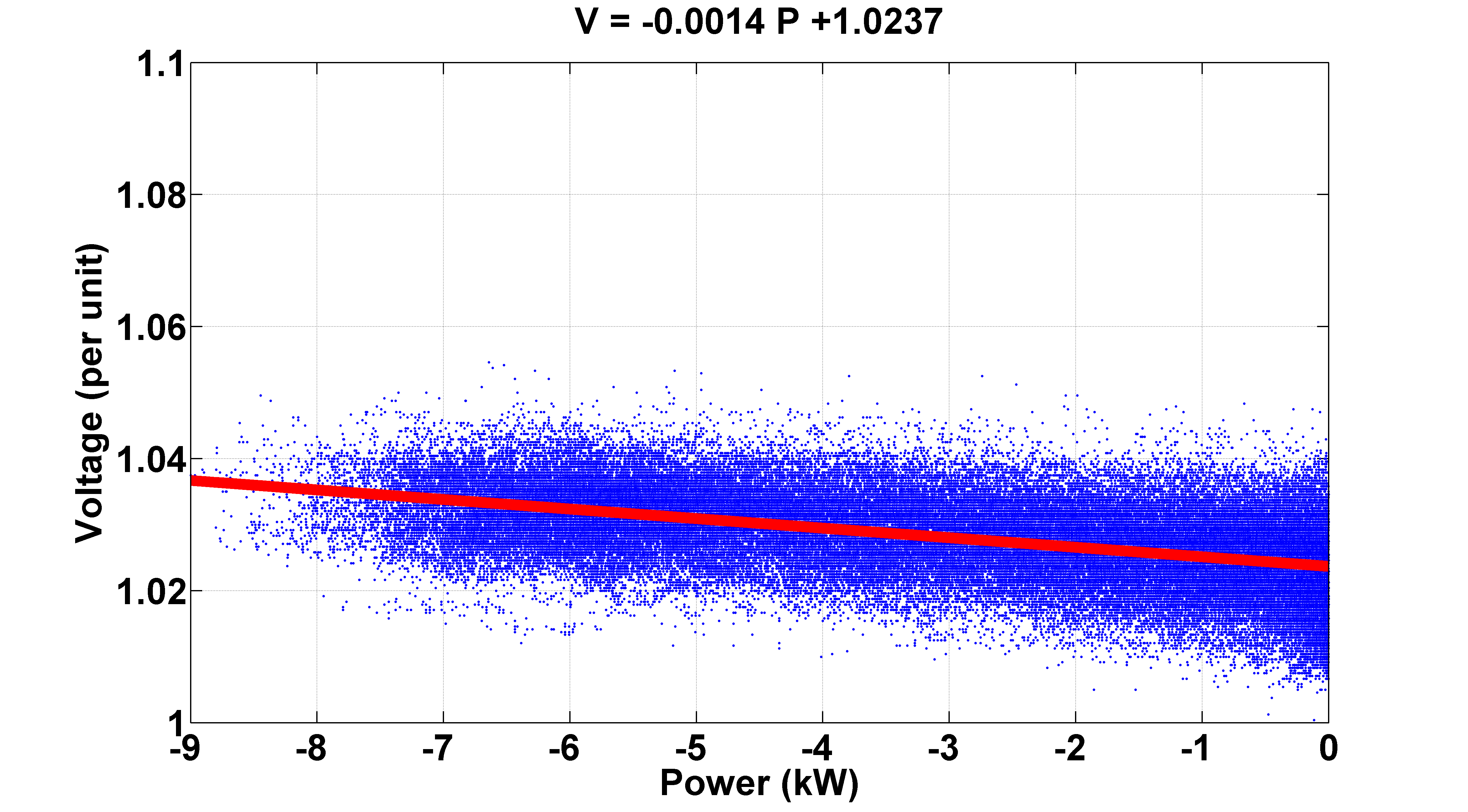}}}
\caption{Voltage versus power for the PV system capacity $9.24$ KW.}
\label{fig:vvp5}

\end{figure}

In this section, we use partial least squares regression, \cite{c3}, to predict the voltage rise caused by the inverter output at the PCC of an inverter. Voltage decreases over distance and load, and in a conventional grid, the delivered voltage at low load is high, in order to ensure adequate voltage at peak load. Keeping the voltage in an acceptable range during low load and peak load states, is a utility challenge. Utilities regulate the voltage using voltage regulators, switched capacitor banks, and transformer load tap changers. The observed voltage at the PCC is affected by the LTC, switched capacitors, neighbours PV systems, and loads. To model the voltage rise caused by the PV system, the effect of LTC and capacitor banks on the voltage must be filtered out. PLS regression is applied to predict the response of voltage to power injected by the PV system. PV system output is considered as the predictor and voltage is considered as the response. The goal of PLS is to predict the voltage variations from injected power. PLS achieves the maximum correlation between power and the voltage, by maximizing the co-variance between power and voltage. PLS combines information about the variances of both the PV system power (predictors) and the voltage (responses), while also considering the correlations between them. Consider $P$ and $V$ as the vectors of measured power and voltage. We seek a scalar $\beta$ such that $V=\beta P +\epsilon$, where $\epsilon$ stands for the error in the model. In figures~\ref{fig:vvp1}-\ref{fig:vvp5}, the voltage versus the power is plotted. The blue dots are the scatter points of the data and the red-lines are the best linear fit to the data using PLS regression. The linear model for the voltage variations versus injected power ($P \leq 0$) for PV system capacities of $1.94$, $3.87$, $7.31$, $11.61$ and $9.24$ KW, are respectively given in (1)-(5).
\begin{flalign}\label{eq:eqno1}
V=-0.0113 P+1.0172, \\ \label{eq:eqno2}
V=-0.0094 P+1.0220, \\ \label{eq:eqno3}
V=-0.0036 P+1.0192, \\ \label{eq:eqno4}
V=-0.0018 P+1.0194, \\ \label{eq:eqno5}
V=-0.0014 P+1.0237.
\end{flalign}
From the comparison of the impedance values in Table~\ref{table:tb0} and equations (1)-(5), it can be concluded that the magnitude of $\beta$ increases with the line impedance. It is observed that the maximum voltage rise for one-kilowatt of injected power is $0.0113$ per voltage unit. Note that the linear model is an approximation to the voltage variation. In the next section more statistics about possible voltage rise are presented.

\subsection{Statistical Parameters}

\begin{figure}[h!]
\centering
\framebox{\parbox{3in}{
\includegraphics[width=3in]{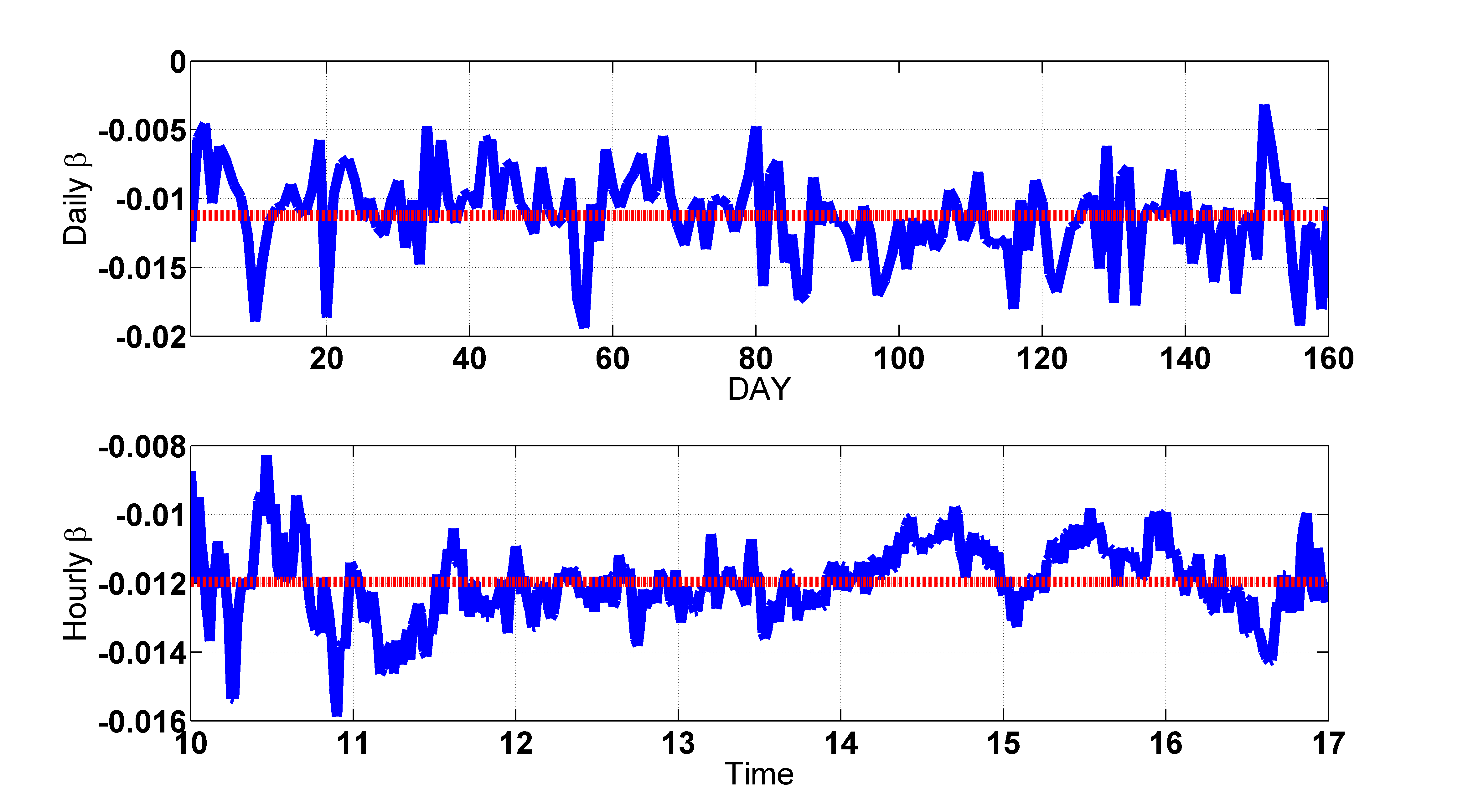}}}
\caption{Daily and hourly variation of $\beta$ for PV system size $1.94$ KW. }
\label{fig:beta1}
\end{figure}

\begin{figure}[h!]
\framebox{\parbox{3in}{
\includegraphics[width=3in]{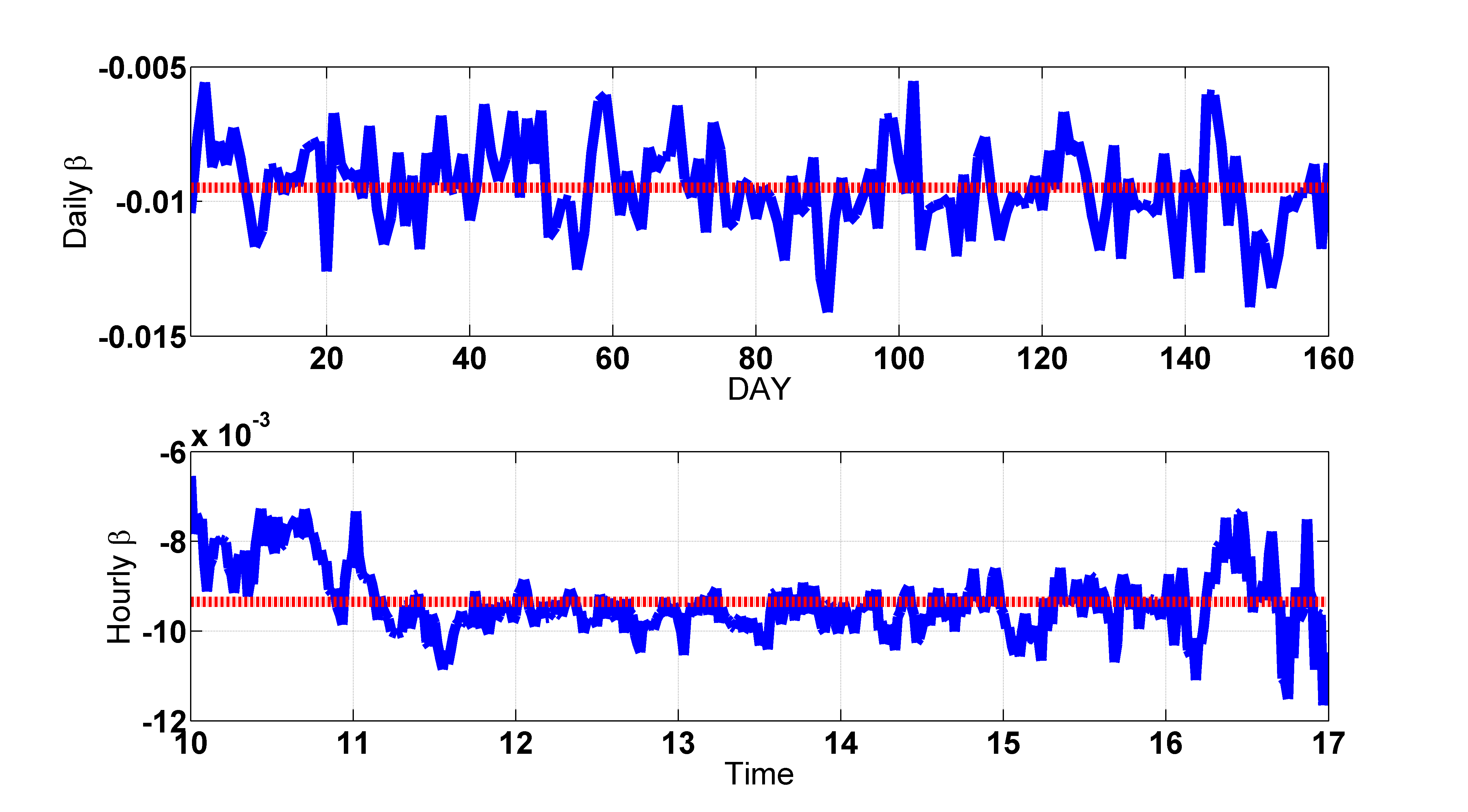}}}
\caption{Daily and hourly variation of $\beta$ for PV system size $3.87$ KW. }
\label{fig:beta2}
\end{figure}

\begin{figure}[h!]
\framebox{\parbox{3in}{
\includegraphics[width=3in]{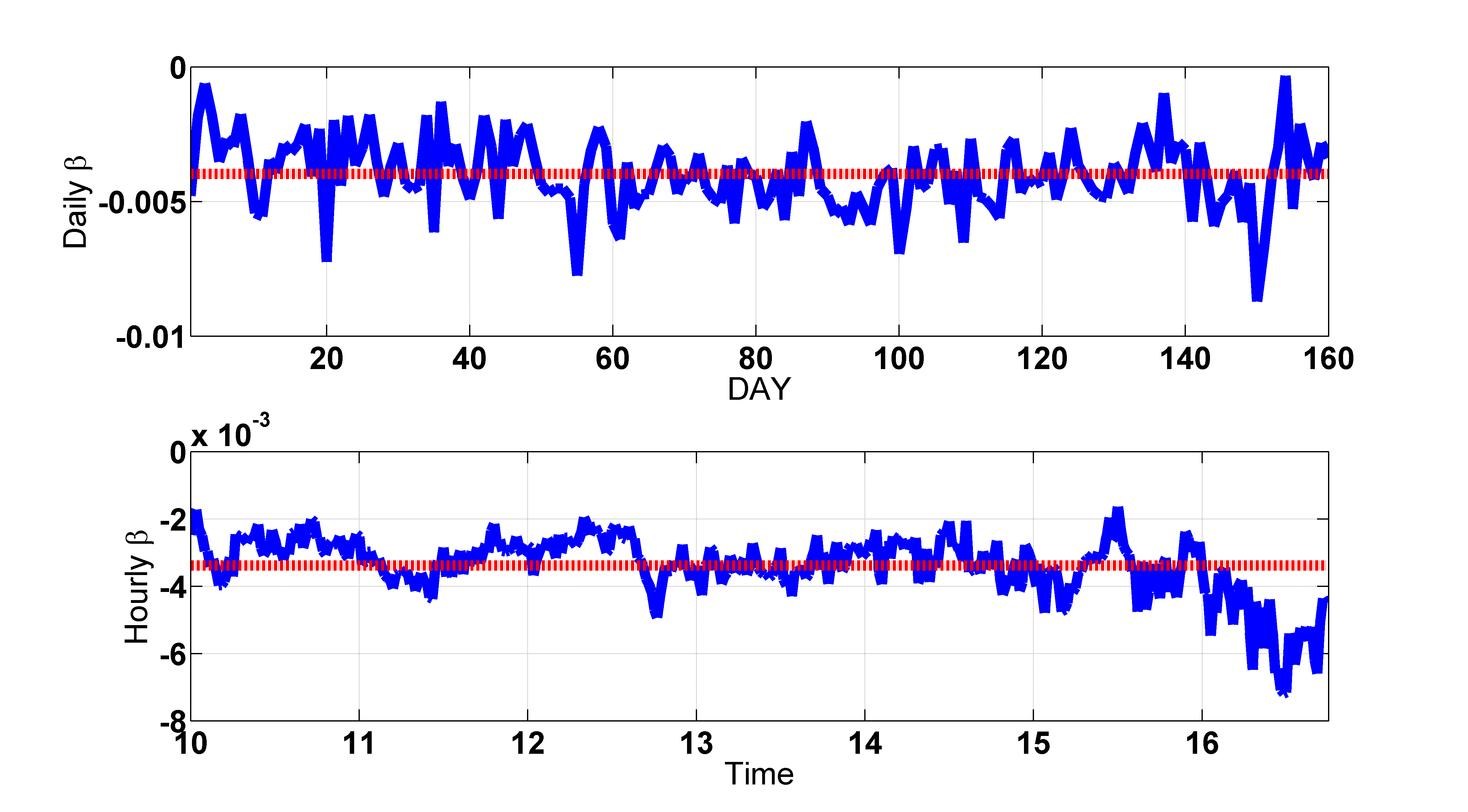}}}
\caption{Daily and hourly variation of $\beta$ for PV system size $7.31$ KW. }
\label{fig:beta3}
\end{figure}

\begin{figure}[h!]
\centering
\framebox{\parbox{3in}{
\includegraphics[width=3in]{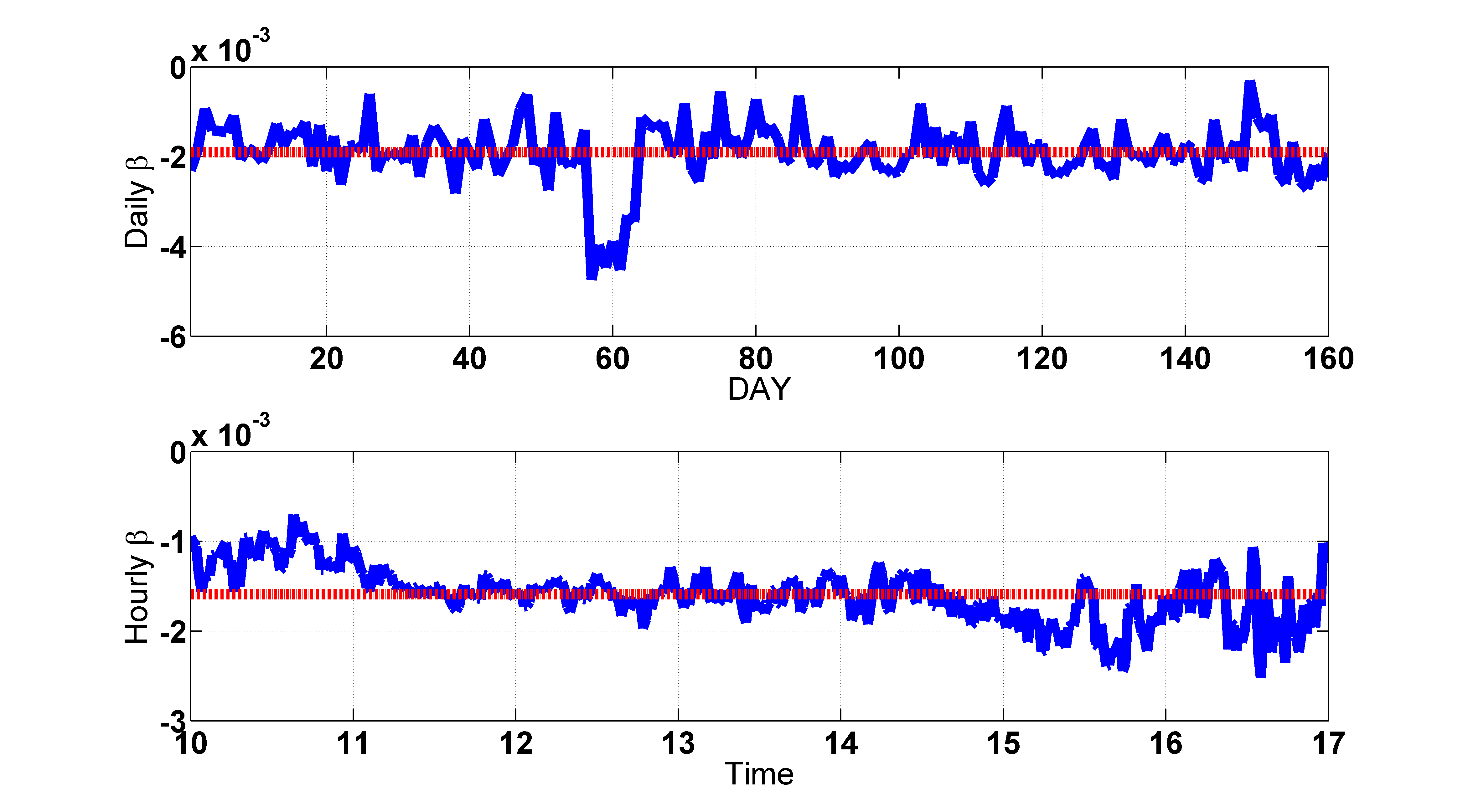}}}
\caption{Daily and hourly variation of $\beta$ for PV system size $11.61$ KW. }
\label{fig:beta4}
\framebox{\parbox{3in}{
\includegraphics[width=3in]{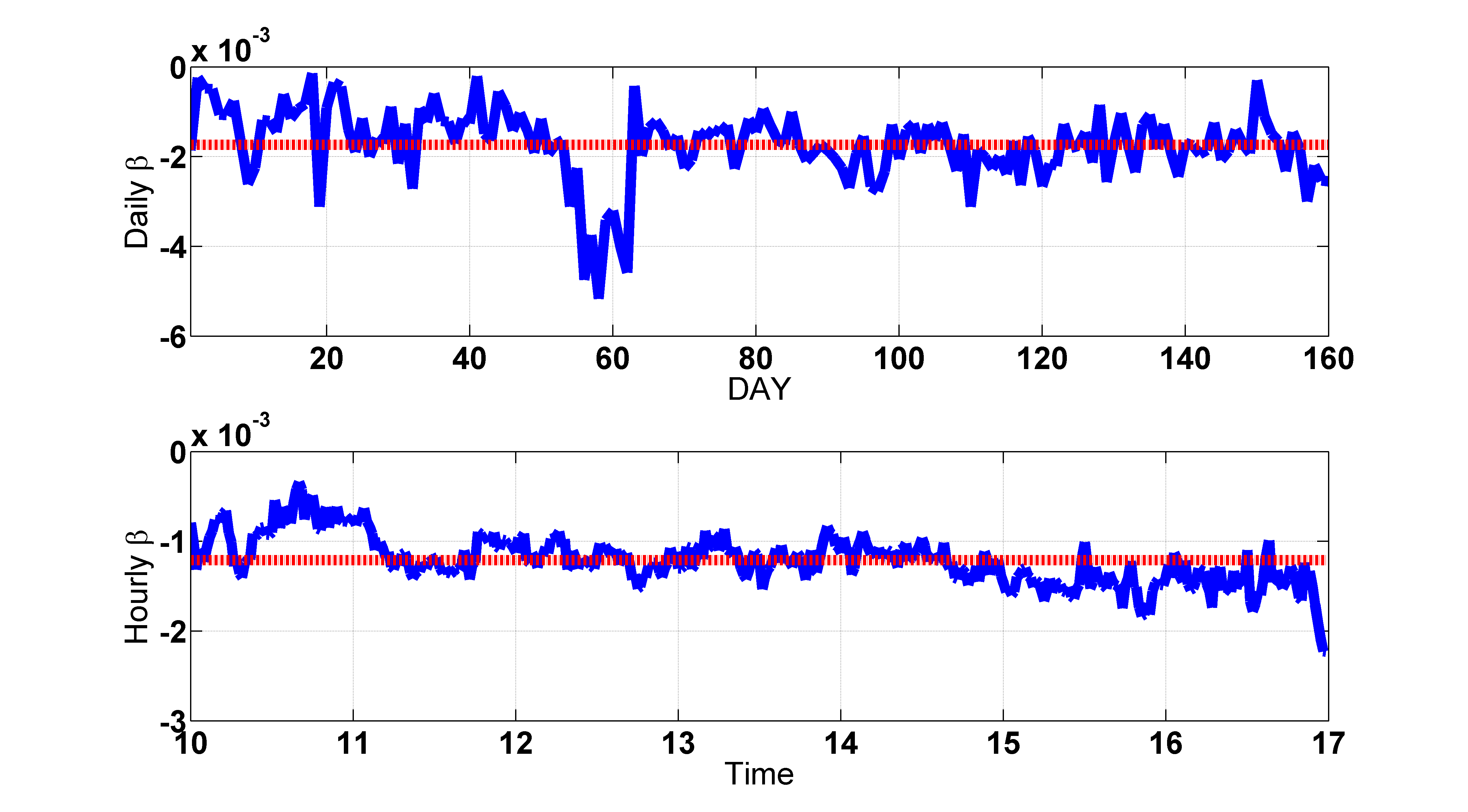}}}
\caption{Daily and hourly variation of $\beta$ for PV system size $9.24$ KW. }
\label{fig:beta5}
\end{figure}

In figures~\ref{fig:beta1}-\ref{fig:beta5}, the daily and hourly variations of $\beta$ are plotted. The dashed red-lines are the averages of $\beta$ in the given figure. In the daily plots, each point corresponds to the value of $\beta$ calculated by applying PLS regression to the corresponding day's data. In the hourly plots, each point corresponds to the value of $\beta$ calculated by applying the PLS regression to the data of the given time over $160$ days. It is observed that the value of $\beta$ has less variation from $12:00$ p.m. to $14:00$ p.m., because of less variation in the loads and stability in the weather conditions.


\begin{figure}[ht!]
\centering
\framebox{\parbox{3in}{
\includegraphics[width=3in]{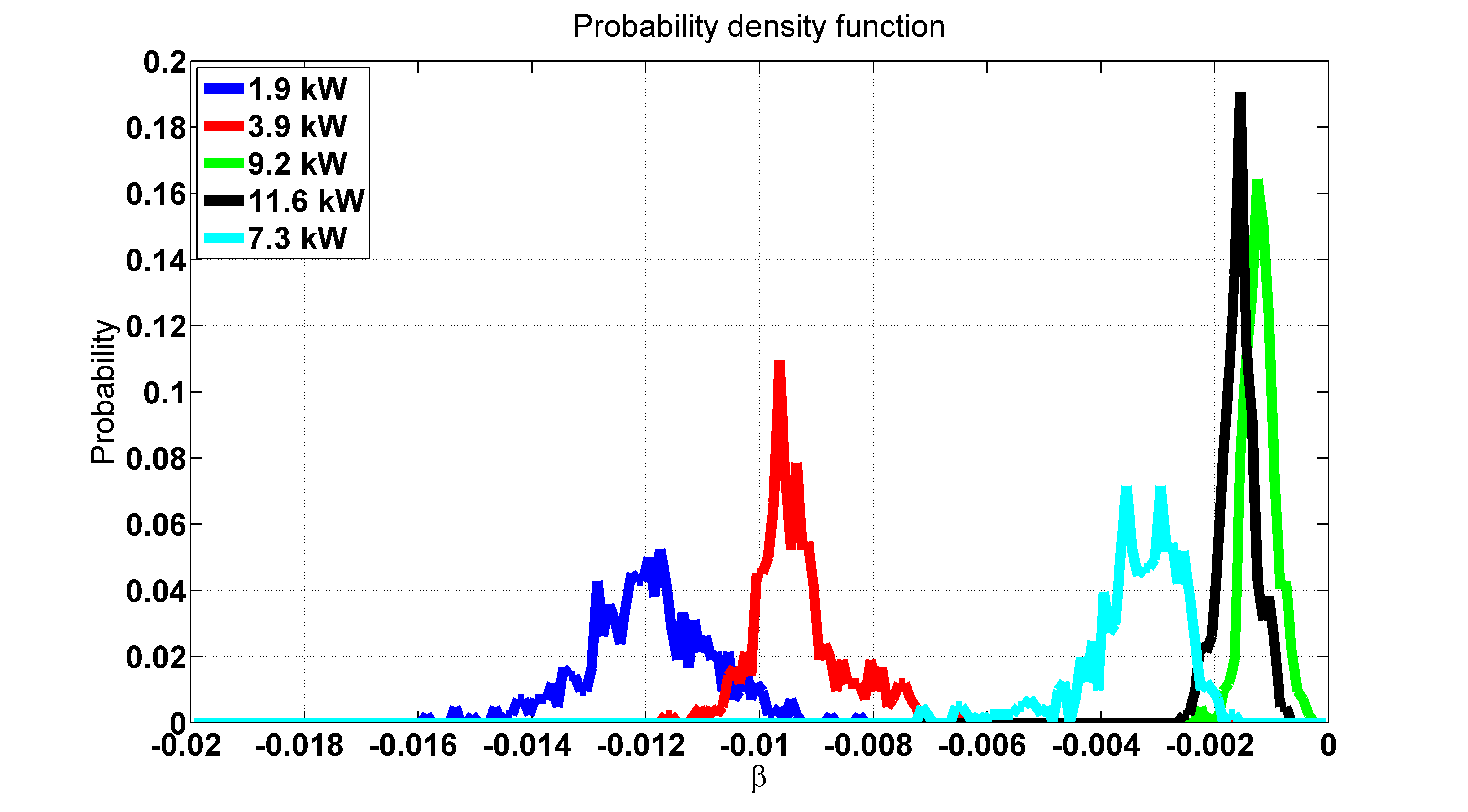}}}
\caption{Daily density for $\beta$.}
\label{fig:density}
\framebox{\parbox{3in}{
\includegraphics[width=3in]{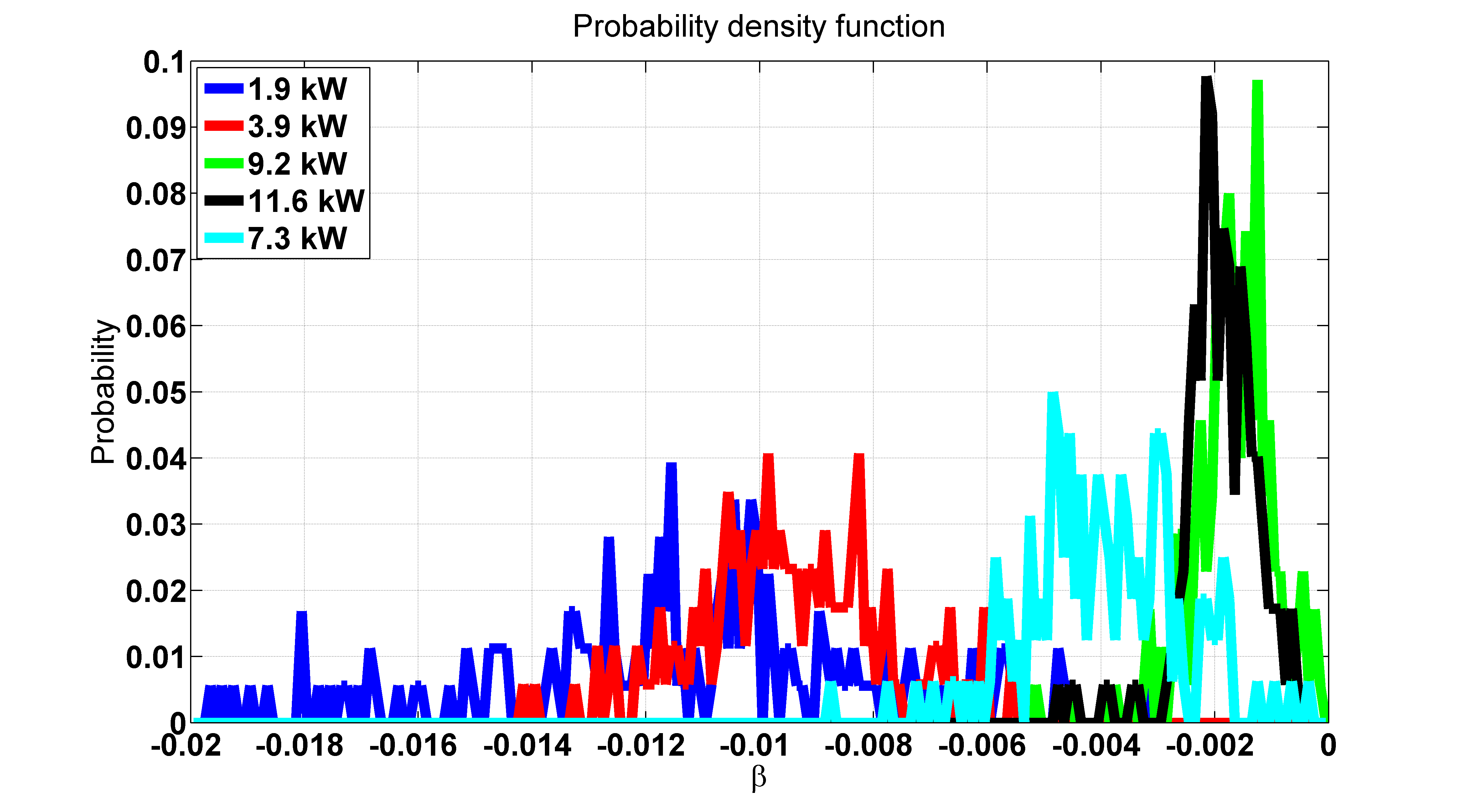}}}
\caption{Hourly density for $\beta$.}
\label{fig:density1}
\end{figure}

In Table~\ref{table:tb2} and \ref{table:tb3} the daily and hourly statistics for the $\beta$ values are presented, i.e., minimum, average, standard deviation (STD), and maximum-a-posteriori (MAP) estimate. By comparison of Table~\ref{table:tb2} and \ref{table:tb3}, it can be deduced that, by injecting one kilowatt power into the grid at locations with PV sizes of $1.94$, $3.87$, $7.31$, $11.61$, and $9.24$ KW, the highest possible voltage increases are respectively $0.0197$, $0.0141$, $0.0087$, $0.0047$ and $0.0052$ per voltage unit. It is observed from Tables~\ref{table:tb0}, \ref{table:tb2} and \ref{table:tb3}, that the magnitude of minimum, average, STD and MAP for $\beta$ increases with line impedance. Therefore, PV systems with larger impedance exhibit more variability in PCC voltage.
\begin{table}[h]
\caption{Daily statistics of $\beta$}
\label{table:tb2}
\begin{center}
\begin{tabular}{|c||c||c||c||c||c|}
\hline
Capacity (KW)& 1.94 & 3.87 & 7.31 & 11.61 & 9.24  \\
\hline
  Minimum   & -0.0197 &  -0.0141 &  -0.0087 &  -0.0047 &  -0.0052   \\
\hline
  Average  & -0.0112 &  -0.0095 &  -0.0040 &  -0.0019 &  -0.0017   \\
\hline
STD    &  0.0034 &   0.0017 &   0.0014 &   0.0008 &   0.0008  \\
\hline
MAP &  -0.0118  & -0.0097 & -0.0036 & -0.0016 &  -0.0013    \\
\hline
\end{tabular}
\end{center}
\end{table}
\begin{table}[h]
\caption{Hourly statistics of $\beta$}
\label{table:tb3}
\begin{center}
\begin{tabular}{|c||c||c||c||c||c|}
\hline
Capacity (KW) & 1.94 & 3.87 & 7.31 & 11.61 & 9.24  \\
\hline
  Minimum  & -0.0159 &  -0.0116 &   -0.0072 &   -0.0025 &  -0.0022   \\
\hline
  Average  & -0.0120 &  -0.0093 &  -0.0034 &  -0.0016 & -0.0012     \\
\hline
STD  &  0.0011 &   0.0009  &  0.0009 &  0.0003  &  0.0003   \\
\hline
MAP & -0.0116  & -0.0099  & -0.0049 &  -0.0022 & -0.0013   \\
\hline
\end{tabular}
\end{center}
\end{table}
In figures~\ref{fig:density} and \ref{fig:density1}, the probability density function for $\beta$ is plotted. In Table~\ref{table:tb4}, the entropies for each of the PCCs is given. It is observed that a PCC with larger impedance has a larger entropy, this implies more randomness and disorder in the behavior of voltage at a PV system with larger line impedance.
\begin{table}[h]
\caption{Entropy of $\beta$}
\label{table:tb4}
\begin{center}
\begin{tabular}{|c||c||c||c||c||c|}
\hline
Capacity (KW)& 1.94 & 3.87 & 7.31 & 11.61 & 9.24  \\
\hline
Daily entropy  & 5.35  &  4.80  &  4.80 &  3.59 &   3.45     \\
\hline
Hourly entropy &  6.31 &   5.77 &  5.39 &   4.5  &  4.5    \\
\hline
\end{tabular}
\end{center}
\end{table}

\section{CONCLUSIONS} \label{SectionVII}

In this work, the effect of injected power into the grid on the voltage fluctuations is quantified. The voltage fluctuations decrease during periods of clear sky, due to less variability in PV generations and loads. Modeling the voltage rise per unit of injected power can be used to develop optimal strategies for operating and placement of voltage regulators. The voltage at each PCC is affected by neighbouring loads and PV systems output. 

\addtolength{\textheight}{-12cm}   

\end{document}